\documentclass[aps,twocolumn,showpacs,amsmath,amssymb,pra,superscriptaddress,floatfix,longbibliography]{revtex4-1}
\usepackage{braket}
\usepackage{amsmath}
\usepackage{amssymb}
\usepackage{mathtools}
\usepackage{graphicx}
\usepackage{dcolumn}
\usepackage{bm}
\usepackage{multirow}
\usepackage{leftidx}
\usepackage{color}
\usepackage{float}
\usepackage{hyperref}

\usepackage{amsfonts}
\usepackage{eufrak}
\usepackage[german,english]{babel}

\newcommand \Hb{$\bar{\rm H}$ }

\begin{document}
\title{Theory of the line shape of the 1S--2S transition for magnetically
trapped antihydrogen}
\author{R. A. Gustafson}
\affiliation{Department of Physics and Astronomy, Purdue University,
West Lafayette, Indiana 47907, USA}
\author{F.~Robicheaux}
\email{robichf@purdue.edu}
\affiliation{Department of Physics and Astronomy, Purdue University,
West Lafayette, Indiana 47907, USA}
\affiliation{Purdue Quantum Science and Engineering Institute, Purdue
University, West Lafayette, Indiana 47907, USA}

\date{\today}

\begin{abstract}
The physics that determines the line shape of the 1S--2S transition in
magnetically trapped $\bar{\rm H}$ is explored. Besides obtaining an
understanding of the line shape, one goal is to replace
the dependence on large scale simulations of $\bar{\rm H}$ with
a simpler integration over well defined functions. For limiting cases,
analytic formulas are obtained. Example calculations are performed
to illustrate the limits of simplifying assumptions. We also describe
a $\chi^2$ method for choosing experimental parameters that can lead to the
most accurate determination of the transition frequency.
\end{abstract}

\maketitle

\section{Introduction}

One of the main goals of the ALPHA collaboration has been to measure the
1S--2S transition in antimatter hydrogen, $\bar{\rm H}$,\cite{ALP2017,ALP2018}
with an accuracy comparable to that
in normal matter H.\cite{PMA2011} The motivation is to compare these two values
as a test of the CPT theorem\cite{AKR2011}; one consequence of the CPT theorem
is that the transition frequencies of $\bar{\rm H}$ and normal hydrogen should
be identical. The frequency in normal matter H is known to an accuracy
of a few Hz.\cite{PMA2011} Currently, this transition frequency is known to an
accuracy of a few kHz in $\bar{\rm H}$\cite{ALP2018}
which is an excellent achievement
considering the few number of $\bar{\rm H}$ in the experiment and
the fact that the transitions occur in a magnetic trap
which shifts the $\bar{\rm H}$ energies.

The transition frequency in the ALPHA experiments is obtained by comparing
the measured line shape to that obtained from a large scale simulation
of the $\bar{\rm H}$ trajectories in the modeled magnetic fields. The
trajectories are needed to understand the positions where the $\bar{\rm H}$s
cross the 243~nm beam and their velocities when crossing. This information
is used to solve for the time dependence of the $\bar{\rm H}$ electronic
states which is used to compute the transition probability for each
crossing. A Monte Carlo sampling of the trajectories then gives the transition
probability and the probability the transition can be detected. This
simulation is a necessary, but somewhat opaque, step in the comparison
of the measured 1S--2S line shape to what is expected assuming CPT.
It is likely that the next generation of experiments will lead to data
giving an accuracy of a few 100 Hz or better.
A few obvious changes will lead to this
improved accuracy: smaller power for the 243~nm laser
to reduce the AC Stark shift, larger 243~nm waist to decrease transit
broadening, colder  $\bar{\rm H}$\cite{DFR2013,ALP2021} to decrease transit
broadening,
and more $\bar{\rm H}$ to decrease the statistical error
bars on the line.

The purpose of this paper is to examine the physics that determines the
line shape of the 1S--2S transition in magnetically trapped $\bar{\rm H}$.
One of the goals is to clarify the role different properties of the
$\bar{\rm H}$ play in the line shape. Another goal is to explain most
aspects of the line shape using analytic formulas that arise in simplified
limits and to show how the large scale simulations approach these
analytic formulas. A final goal is to explore a $\chi^2$ method for
predicting how choices for experimental parameters affect the uncertainty
in the frequency measurement.

This paper is organized as: Sec.~\ref{sec:Bas} contains a
description of the basic physics determining the line shape,
Sec.~\ref{sec:Trans1} contains a description of how to calculate
the transition probability for one $\bar{\rm H}$ crossing of the
243~nm beam, Secs.~\ref{sec:LiShnoB} (\ref{sec:LiShB}) contain
descriptions for
the line shape when the shifts of the frequency due to a change
in the magnetic field are not (are) included, 
Sec.~\ref{sec:OptPa} contains a description of a method for using $\chi^2$ to
choose the experimental parameters that would give the most accurate
frequency determination, and
Sec.~\ref{sec:Sum} contains a summary of the results.

%

\section{Basic physics of the 1S--2S transition}\label{sec:Bas}

\subsection{Excitation}

The two photon absorption from counter-propagating laser beams gives 
a transition from the 1S to the 2S state with no first order Doppler
shift. Because the lifetime of the 2S state in zero electric and
magnetic fields is $1/8.2$~s,
external factors (e.g. laser waist and power, $\bar{\rm H}$ temperature,
magnetic fields, etc) mainly determine the line shape of the
1S--2S transition. This section sketches how to incorporate these
effects.

We assume that a Gaussian beam well approximates the light in the trap.
If this assumption is violated, most of the analytic results below will
no longer be accurate, but the results from solving the optical Bloch
equations can incorporate different laser shapes.
For a description of a Gaussian
beam, we will take the $z$ direction to be along the beam with the $x,y$
directions perpendicular to the beam. The intensity at the center of one
beam is $I_0=2P_0/(\pi w^2_0)$ where $P_0$ is the power in the
beam and $w_0$ is the beam waist.
The position dependent intensity for a single Gaussian beam
is
\begin{eqnarray}
I(r,z) &=& I_0 [w_0/w(z)]^2\exp [-2r^2/w^2(z)]\label{eq:Igau}\\
w^2(z)&=&w_0^2 (1+z^2/Z_R^2)\label{eq:wgau}
\end{eqnarray}
where $r^2=x^2+y^2$ and the Rayleigh range
$Z_R=\pi w_0^2/\lambda$.
For the 1S--2S transition, $\lambda = 243$~nm.
For counterpropagating beams, the electric field at a position
$r,z$ is $E=E_0 [w_0/w(z)]\exp [-r^2/w^2(z)]\cos [\phi (r,z)+\delta]
\cos(\omega_L t)$ which has the form of a standing wave; the spatial phase
dependence, $\phi (r,z)$, is not relevant for our results. The relationship
between $E_0$ and the one beam maximum intensity is
$E_0^2=8I_0/(\varepsilon_0 c)$.

The theory for this transition has been discussed in several
places; we will follow the treatment in Ref.~\cite{RMR2017}. The
coupling of the 1S and 2S states proceeds through a virtual transition
to the bound $nP$ and continuum $\mathcal{E} p$-states. After adiabatically
eliminating the $p$-states, the equations governing the two-photon coupling
between the 1S and 2S states are
\begin{eqnarray}
\frac{dC_{1S}}{dt}&=\frac{\xi E_0^2w_0^2}{i\hbar w^2}e^{-2r^2(t)/w^2} e^{-i(\mathcal{E}_{2S}-\mathcal{E}_{1S}-2\hbar\omega_L )t/\hbar} C_{2S}\label{eq:C1SC2S1}\\
\frac{dC_{2S}}{dt}&=\frac{\xi E_0^2w_0^2}{i\hbar w^2}e^{-2r^2(t)/w^2} e^{i(\mathcal{E}_{2S}-\mathcal{E}_{1S}-2\hbar\omega_L )t/\hbar} C_{1S} \label{eq:C1SC2S}
\end{eqnarray}
where $\omega_L=2\pi f_L$ is the angular frequency of the laser with
$f_L$ the laser frequency. and
$\mathcal{E}_{2S}-\mathcal{E}_{1S}$ is the energy difference between
the 1S and 2S states at the position of the crossing. We will use $f$ in all
expressions for frequency. Instead of computing
$\xi$ by summing over the infinite number of $nP$ states and integrating
over the continuum $\mathcal{E} p$-states, we perform the calculation with
the atom inside a spherical
box so that the number of negative energy states is finite and the continuum
is discretized. If the radius of the box is sufficiently large, $\xi$ is
independent of the value of the radius.
The parameter $\xi$ is
\begin{equation}
\xi =-\frac{e^2}{8}\sum_n\frac{D_{2S,nP}D_{nP,1S}}{\mathcal{E}_{nP}-\mathcal{E}_{1S}-\hbar\omega_L} \simeq 12.3\varepsilon_0 a_0^3,
\end{equation}
where $a_0$ is the Bohr radius,
$D_{2S,nP}D_{nP,1S}=\sum_m\langle 2S|\vec{r}|nPm\rangle\cdot
\langle nPm|\vec{r}|1S\rangle $, with $m$ the azimuthal quantum
number, and $e$ is the electric charge. 
The numerical value was obtained by performing the sum using
$nP$ states whose radial wave function is zero at 30$\,a_0$.

There are several effects that are missing from these equations which
will be added or discussed below. The main missing effects are:
the AC Stark shift which arises because the 1S and 2S states have
different AC polarizabilities, the second order Doppler
shift proportional to the kinetic energy over the rest energy
of the $\bar{\rm H}$, ionization of the 2S state by a third photon,
radiative decay from the 2S state, mixing of the 2S and 2P states
due to the $v\times B$ effective electric field, etc.

The transition frequency depends on the spin coupling of the positron
and antiproton and the magnetic field. The 1Sc,2Sc states have total angular
momentum 0 in the $B$-field direction while the 1Sd,2Sd states have the two
spins antiparallel to the $B$-field direction. Because the 1Sc--2Sc
transition has a $\sim 10\times$ larger variation with $B$ when
$B\sim 1$~T, we will restrict the examples to the 1Sd--2Sd transition.
The change in frequency with $B$ is given by\cite{RMR2017}
\begin{equation}\label{eq:dEdB}
\frac{d\;\null}{dB}\frac{\mathcal{E}_{2S}-\mathcal{E}_{1S}}{2h} =  (\frac{1}{2}186.071 + 387.678\, B)\; {\rm kHz}
\end{equation}
for $B$ in Tesla.
The second term is from the diamagnetic term in the Hamiltonian and
causes, at larger $B$, a larger variation of the transition frequency
with $B$.

Another shift can occur due to a $v\times B$ effective electric
field causing an interaction of the 2S with the 2P states. Using
Eq.~(43) of Ref.~\cite{RMR2017}, this shift is
\begin{equation}
\Delta {\cal E}_{2S}/h \sim 0.041\; v_\perp^2\; {\rm Hz}
\end{equation}
when $B\sim 1$~T and the perpendicular velocity, $v_\perp$, is in $m/s$.
For a $\bar{\rm H}$ with a perpendicular kinetic energy of
50~mK, this shift corresponds to $\sim 40$~Hz which is negligible
for the next level of accuracy in ALPHA experiments. This shift can be
decreased by cooling the $\bar{\rm H}$s.

\subsection{Detection}

There are several processes that lead to transitions out of the
2S state which can be used to detect the excitation. Ordinary matter
experiments\cite{PMA2011} detect photons emitted after excitation to the
2S state.
Detecting emitted photons is probably
unfeasible for $\bar{\rm H}$ which is trapped in a long tube.
Therefore, other processes are important for detection
of the 1S--2S transition in the ALPHA experiments.

Two processes
are presumed to dominate the detection in experiments reported
previously.\cite{ALP2017,ALP2018} The first is ionization of the 2S state
by a third 243~nm photon. This can occur during the excitation process
itself or when an excited $\bar{\rm H}$ recrosses the 243~nm beam
at a later time. The second is when the $v\times B$ effective electric
field causes mixing with 2P states where the positron has the untrapped
spin orientation. This allows a one photon emission back to the ground
state into magnetically untrapped 1S states. Both of these processes
result in annihilation on the trap wall as the detection step.
Unfortunately, both
depend on the perpendicular speed of the $\bar{\rm H}$ to some extent
which affects the measurement of the transition line shape.
However, we will argue below that future experiments should use substantially
lower 243~nm laser power and colder $\bar{\rm H}$s to achieve higher
precision in the frequency measurement. In this case, neither of these
mechanisms will be effective: the ionization is proportional to the laser
power and the spin flip is proportional to the temperature.

One possibility is to impose a weak electric field
which would cause mixing between the
2S and 2P states. This can lead to a spin flip after a one photon decay
back to the ground state. If the electric field were larger than
$|v\times B|$, then the decay would be relatively independent of the
$\bar{\rm H}$ position and velocity distribution. Another possibility
is to
stimulate transitions from the 2S
to the 2Pf state using microwaves.
The 2Pf state has a large probability to decay to untrapped 1S states
which lead to annihilation on the trap walls.
Microwave intensity of 
$\sim 0.01-0.1$~mW/cm$^2$ would be sufficient to make it the
dominant decay process. More importantly, the transition rate will be nearly
independent of the $\bar{\rm H}$ velocity and
position distribution. Only when the $\bar{\rm H}$
travels to regions of higher $B$-field will the transition rate change
because the frequency of the microwave transition depends on $B$.
However, most of the $\bar{\rm H}$ trap has nearly the same $B$-field
which is why this transition will not strongly depend on the $\bar{\rm H}$
distribution. Thus, a benefit of both detection methods is that the
transition line shape is not distorted by the detection process.

\section{Transition probability for one beam crossing}\label{sec:Trans1}

In this section, we give the expressions for the probability for
a transition into the 2S state when the $\bar{\rm H}$ crosses
a Gaussian beam of intensity $I(z)$.
Because the beam has a finite width, there is a finite time for the
$\bar{\rm H}$ to cross the beam, leading to a line width
(transit broadening) roughly the inverse
of the time to cross the beam. The material in this section briefly
summarizes the derivation in Ref.~\cite{RMR2017}.

\subsection{Perturbative expression}

When the laser is weak enough, saturation of the transition, the AC Stark shift,
and ionization
out of the 2S state are negligible effects. If the atom crosses
the laser beam quickly enough, the radiative decay of the 2S state
can also be neglected. In this case, setting $C_{1S}=1$ in
Eq.~\eqref{eq:C1SC2S} is a good approximation and an integral over
time will give the amplitude to transition to the 2S state.

We are interested in the case where the beam waist is much smaller than
the scale over which the magnetic and electric fields vary substantially.
This will lead to a position dependent detuning which we define through
$\hbar\Delta \equiv 2\hbar\omega_L - (\mathcal{E}_{2S}-\mathcal{E}_{1S})$
with the 1S and 2S energies evaluated at the point where the 
$\bar{\rm H}$ crosses the beam. Given these conditions, the $\bar{\rm H}$
will have nearly constant velocity so that $r^2(t)=b^2+v_\perp^2t^2$
with $b$ the distance of closest approach to the beam axis and $v_\perp$
the magnitude of the velocity perpendicular to the beam axis. The
resulting integral is the Fourier transform of a Gaussian which leads
to the probability for a transition:
\begin{equation}\label{eq:2sprob}
|C_{2S}|^2 \simeq 32 \pi  \frac{\xi^2 I_0^2}{\hbar^2 \varepsilon_0^2c^2} \frac{w_0^4}{w^2v_\perp^2} e^{-4b^2/w^2} e^{-[2\pi w\Delta f/v_\perp ]^2}
\end{equation}
where $\Delta f=f_L-f_0$ with $f_0=(\mathcal{E}_{2S}-\mathcal{E}_{1S})/(2h)$ 
and the waist, $w$,
evaluated at the distance of closest approach. In all expressions,
the $z$ dependence of the intensity and waist, $I(z),w(z)$ in
Eqs.~\eqref{eq:Igau} and \eqref{eq:wgau}, will
not be explicitly written for notational convenience. If the laser has
a substantial linewidth, this expression needs to be convolved with 
the frequency distribution of the laser as in Eq.~(32) of Ref.~\cite{RMR2017}.
We will assume this is a small fraction of the width due to the finite
crossing time and not include the linewidth of the laser below.

\subsection{Optical Bloch equation}

\begin{figure}
\resizebox{80mm}{!}{\includegraphics{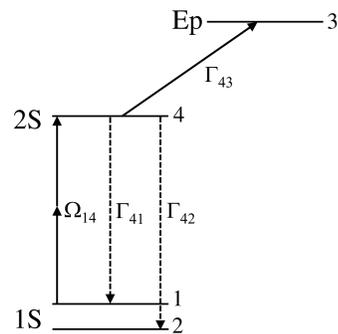}}
\caption{\label{fig:OBschem}
Schematic of the excitation out of the trappable 1S-state $|1\rangle$
into the trappable 2S-state $|4\rangle$. The 2S-state population can
decay by photon emission into the 1S-states and by absorption of a third
243~nm photon into the $p$-continuum, $|3\rangle$.
}
\end{figure}

In the previous section, we made several assumptions that could affect
the line shape. This section will give the equations that can be solved
for a more accurate calculation of the transition probability. We follow
the derivation of Ref.~\cite{RMR2017} by using the density matrix
formalism to describe the evolution of the electronic states of the
$\bar{\rm H}$, Fig.~\ref{fig:OBschem}.
We only include 4 states in this treatment: $|1\rangle$
is the low field (trappable) 1S state which initially has 100\%
of the population, $|2\rangle$ is a high field (untrappable) 1S state
which can be produced in decays from the 2S state, $|3\rangle$ represents
photo-ionization which results when the 2S state absorbs a third
photon, and $|4\rangle$ is the low field (trappable) 2S state which
is reached in the two photon transition from the 1S state.
Properly speaking, $|3\rangle$ is not a state but a continuum of states,
Ep;
approximating photoionization as decay to a single state
can be used because we are only interested in the
total population of ionized atoms and it only enters the density matrix
equation through decay terms.

The density matrix equations are written in Lindblad form
\begin{equation}
\frac{d\hat{\rho}}{dt}=\frac{1}{i\hbar}[\hat{H},\hat{\rho}]+\mathcal{L}(\hat{\rho})
\end{equation}
which leads to the following equations for the non-zero
matrix elements of $\hat{\rho}$:
\begin{eqnarray}\label{eq:OB}
\dot{\rho}_{11} &=& - \frac{i}{2} \Omega_{14}(t) \left( \rho_{41} - \rho_{14} \right) + \Gamma_{41} \rho_{44}\nonumber\\ 
\dot{\rho}_{22} &= &\Gamma_{42} \rho_{44} \nonumber\\ 
\dot{\rho}_{33} &= &\Gamma_{43} \rho_{44} \\ 
\dot{\rho}_{44} &=& - \frac{i}{2} \Omega_{14}(t) \left( \rho_{14} - \rho_{41} \right) - \Gamma \rho_{44}\nonumber\\ 
\dot{\rho}_{14} &=& - \frac{i}{2} \Omega_{14}(t) \left( \rho_{44} - \rho_{11} \right) + \left( -i \Delta_{AC} - \frac{1}{2} \Gamma \right)\rho_{14}\nonumber
\end{eqnarray}
where $\Delta_{AC}=\Delta -2\pi\Delta f_{\rm AC}(t)$ and
$\rho_{41}=\rho_{14}^*$ determines the last non-zero element. The
$\Gamma_{4i}$ are the decay rates to the different final states,
$\Gamma$ is the sum of these rates, and $\Omega_{14}$ is the two-photon
Rabi frequency.
The two beam
AC Stark shift frequency for the 1S--2S transition is
\begin{equation}\label{eq:ACst}
\Delta f_{\rm AC}(t) =2I\textup{e}^{-2r^2(t)/w^2}1.67\; {\rm Hz}
\end{equation}
where $I$ is the intensity of one beam at $z$ and $r = 0$ in ${\rm W/cm^2}$
and the factor
$1.67$~Hz is from Ref.~\cite{HJK2006}.
For two counterpropagating 1~W beams with 200~$\mu$m waist,
the $\Delta f_{\rm AC}\approx 5$~kHz.
The
two photon Rabi frequency is
\begin{equation}
\Omega_{14}(t)= \frac{16\xi I}{\hbar\varepsilon_0 c} \textup{e}^{-2r^2(t)/w^2}.
\end{equation}
The total decay rate of the 2S state is $\Gamma = \Gamma_{41}+\Gamma_{42}+
\Gamma_{43}$ where $\Gamma_{41},\Gamma_{42}$ is the radiative decay rate
into the trapable and untrapable 1S states, respectively; see
Ref.~\cite{RMR2017}
for these decay rates. If a microwave or static electric
field is causing transitions
from the 2S to the 2P states, then the $\Gamma_{41},\Gamma_{42}$ will
be increased by factors depending on the strength and detuning
of the microwaves or the strength and direction of the electric
field. Finally,
the ionization rate out of the 2S state is
\begin{equation}
\Gamma_{43}= I
\textup{e}^{-2r^2(t)/w^2} 7.57 \; \textup{s}^{-1}\label{eq:Gamma_ion}
\end{equation}
where $I$ is in ${\rm W/cm^2}$ and the 7.57 was determined by numerically solving
for the photo-ionization cross section out of the 2S state from
243~nm photons. For two counterpropagating 1~W beams with
200~$\mu$m waist, the $\Gamma_{43}\approx 2\pi\; 4$~kHz.

In the calculations below, we numerically solve the density matrix equations
using $I(t)$ and $r(t)$ for individual atom trajectories.

\section{Line shape: no magnetic or electric fields}\label{sec:LiShnoB}

In this section, we give results when the shifts in transition frequency
due to external $E$- and $B$-fields are ignored. The perturbative
transition rate can be analytically calculated for a thermal
distribution of $\bar{\rm H}$ velocity as well as for an equal
distribution of velocities within a sphere in velocity space.
The perturbative transition rate
can be reduced to a single integral when the distribution
only depends on the $\bar{\rm H}$ kinetic energy.

\subsection{Perturbative result}

For this section, we use the perturbative calculation of the transition
probability for one beam crossing, Eq.~\eqref{eq:2sprob}, as the starting
point. We then average over the $\bar{\rm H}$ positions and
velocities to get the rate for transition
into the 2S state. To simplify the notation, we will combine the terms
in the probability that do not contain $v_\perp$ or $b$ into
\begin{equation}
A\equiv 32 \pi \frac{\xi^2 I_0^2}{\hbar^2\varepsilon_0^2 c^2}\, \frac{w_0^4}{w^2}.
\end{equation}

The rate that $\bar{\rm H}$s pass the beam with a distance between $b$ and
$b+db$ is
\begin{equation}
{\cal F} =\rho_{2D}v_\perp 2 db
\end{equation}
where the $\rho_{2D}$ is the two-dimensional $\bar{\rm H}$ density and
the 2 arises because the $\bar{\rm H}$ can pass on either side of the
beam for a given direction $\hat{v}_\perp$. The probability distribution
for finding an $\bar{\rm H}$ with a perpendicular speed between $v_\perp$
and $v_\perp + dv_\perp$ will be called $v_\perp {\cal D}(v_\perp )dv_\perp$.

The rate of $\bar{\rm H}$s transitioning from the 1S to the 2S state divided
by the two-dimensional $\bar{\rm H}$ density is
\begin{eqnarray}\label{eq:avgrate}
{\cal G}&=&2A\int_0^\infty e^{-4b^2/w^2}db\int_0^\infty
{\cal D}(v_\perp )e^{-[2\pi w\Delta f/v_\perp ]^2}
dv_\perp\nonumber\\
&=&\frac{Aw\sqrt{\pi}}{2}\int_0^\infty
{\cal D}(v_\perp )e^{-[2\pi w\Delta f/v_\perp ]^2}dv_\perp
\end{eqnarray}
Note that the ${\cal G}$ has units of $area/time$.

We note that this expression is somewhat problematic for small
$\Delta f$ because the
perturbation calculation of the transition probability has a
factor of $1/v_\perp^2$ which can cause $|C_{2S}|^2$ to be larger
than 1 for small $\Delta f$ which is
impossible. Thus, the perturbative calculation of the line shape will
be inaccurate for small detuning. The range of detuning where the
line shape is inaccurate decreases as the intensity decreases.

\subsubsection{Thermal distribution}

The results in this section reproduce those in Ref.~\cite{BBC1979} for
the special case of equal intensity in the counter-propagating beams.
The distribution of $v_\perp$ for a thermal distribution is
${\cal D}_{th}=(2 /v_{th}^2)\exp (-v_\perp^2/v_{th}^2)$ with
$v_{th}^2\equiv 2k_BT/M$ with $T$ the temperature and $M$ the
mass of the $\bar{\rm H}$. The thermal transition rate into the
2S state is
\begin{eqnarray}\label{eq:thrate}
{\cal G}_{th}&=&\frac{Aw\sqrt{\pi}}{v_{th}^2}\int_0^\infty \exp \left[
-\frac{v_\perp^2}{v_{th}^2}-\frac{(2\pi w \Delta f)^2}{v_\perp^2}
\right] dv_\perp\nonumber\\
&=&\frac{\pi Aw}{2v_{th}}e^{-|f_L-f_0|/\phi}
\end{eqnarray}
where $\phi = v_{th}/(4\pi w)$
which gives a linewidth proportional to $v_{th}/w$ as expected (although
the exponential of the absolute value of the detuning is an interesting
functional form).

Because the perturbative calculation is problematic for small detuning,
we expect the transition rate to be modified for $f_L\simeq f_0$.
Therefore, the discontinuous change in slope of ${\cal G}_{th}(f_L)$
with respect to the laser frequency will be modified for the actual
transition.

When the $\bar{\rm H}$s are in a trap, the $v_\perp$ distribution,
${\cal D}(v_\perp )$, must exactly go to zero for energies that can
escape the trap.
Therefore, the thermal distribution will only be
relevant for $k_B T$ much less than the trap energy. For the reported ALPHA
experimental results,
the $\bar{\rm H}$ trap depth is $E/k_B\sim 1/2$~K.

\subsubsection{Energy dependent distribution}\label{sec:EnDepVel}

The next case  we consider is when the $v_\perp$ distribution arises
from a distribution with respect to energy in 3D with the velocity in the
$z$-direction averaged out. In this case, the ${\cal D}(v_\perp )$ will
be a function of $v_\perp^2$. For a general case, the integral in
Eq.~\eqref{eq:avgrate} needs to be performed numerically. Although
the essential singularity at $v_\perp =0$ looks bad, the integrals
over one parameter can be evaluated by simply increasing the number of
points.

As an example that can be done analytically,
consider the early ALPHA experiments where the distribution of $\bar{\rm H}$s
could be considered as the low energy portion of a high temperature
distribution. As an extreme example, we consider the case where the
velocity distribution is flat in $v_x,v_y,v_z$ up to the condition
$v_m^2>v_x^2+v_y^2+v_z^2$; this is a flat distribution within a sphere
in velocity space. This gives ${\cal D}_m=3\sqrt{v_m^2-v_\perp^2}/v_m^3$.
Using this distribution, the transition rate in units of $area/time$ is
\begin{eqnarray}\label{eq:hirate}
{\cal G}_m&=&\frac{Aw\sqrt{\pi}3}{2v_m}\int_0^1
\sqrt{1-s^2}\; e^{ -\eta^2 /s^2}ds\nonumber\\
&=& \frac{3Aw\pi\sqrt{\pi}}{8v_m}\left[(2\eta^2+1){\rm erfc} (|\eta |)-
\frac{2|\eta |}{\sqrt{\pi}}e^{-\eta^2}\right]
\end{eqnarray}
where $\eta =2\pi w (f_L-f_0)/v_m$.
As with the previous section, this expression will be least accurate
for $f_L\simeq f_0$ but will become accurate over a larger range
as the laser intensity decreases.

\begin{figure}
\resizebox{86mm}{!}{\includegraphics{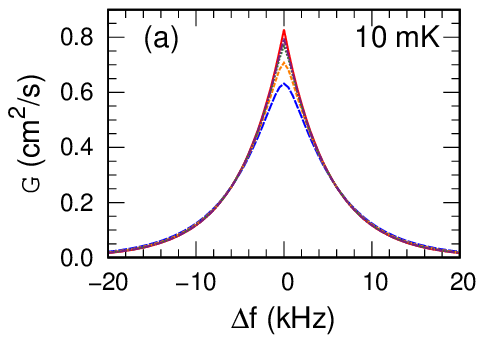}\includegraphics{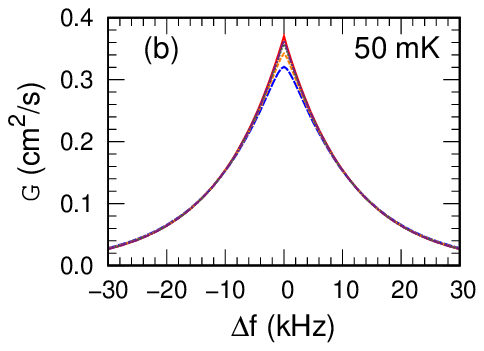}}
\resizebox{86mm}{!}{\includegraphics{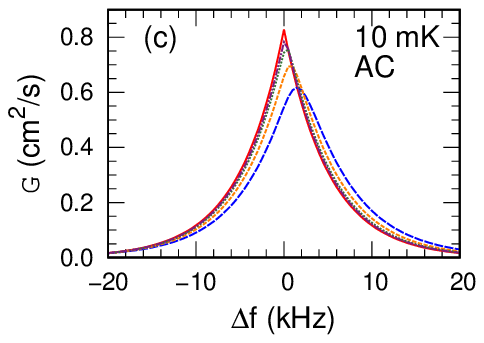}\includegraphics{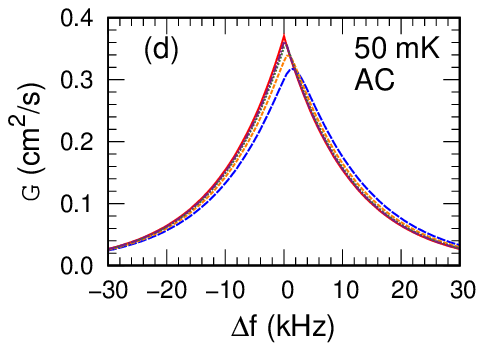}}
\caption{\label{fig:OBth}
The thermally averaged 1S--2S transition rate. All
results use a beam waist of 200~$\mu$m and don't include magnetic
and electric field shifts versus detuning of the two
photon transition. The results are all scaled by a
factor of $1/P_0^2$ with $P_0$ the power in one beam so that the
lines are on the same scale for different powers.
The perturbative treatment, Eq.~\eqref{eq:thrate},
(red solid line) does not change with intensity. The optical
Bloch results change with power: 1~W (blue dashed), 0.5 W (orange
short dash), 0.2 W (green dotted), and 0.1 W (purple dash-dot).
(a) and (c) are for 10~mK thermal distribution (b) and (d) are for 50~mK thermal distribution. Plots (c) and (d) include the AC Stark shift in the
optical Bloch equations while (a) and (b) do not.
}
\end{figure}

\subsection{Optical Bloch result}\label{sec:OB1}

Because this section only investigates the excitation of the
2S state, we will set the branching ratio of the radiative
decay to be 100\%
into the untrapped 1S state. With this condition, the excitation
rate divided by the two-dimensional $\bar{\rm H}$ density is
\begin{equation}\label{eq:OBrate}
{\cal G}_{ob}=\int_0^\infty\int_0^\infty 2v_\perp^2 {\cal D}(v_\perp)
(1-\rho_{11})db dv_\perp
\end{equation}
where the density matrix element, $\rho_{11}$
is evaluated at large time for parameters $b$ and $v_\perp$.

\subsubsection{Thermal distribution}

In this section, we present
results from numerically solving the optical Bloch
equations and using the result to calculate the rate, ${\cal G}$.
We solved the optical Bloch equations, Eq.~\eqref{eq:OB}, using
equal steps in $db$ to sample the crossing distance, $b$, and
equal steps in $dv_\perp$ to sample
the perpendicular speeds, $v_\perp$. From above, the thermal
distribution gives ${\cal D}(v_\perp )=(2/v_{th}^2)
\exp (-v_\perp^2/v_{th}^2)$ in Eq.~\eqref{eq:OBrate}.
We performed calculations
for two temperatures, 10 and 50~mK, and laser powers from 0.1
to 1~W to illustrate the limitation of the perturbative line
shape, Eq.~\eqref{eq:thrate}.

In order to more easily compare the results for different laser powers,
we scaled the ${\cal G}$ by dividing by the squared laser power.
The calculations were done for a 200~$\mu$m waist and do not include
shifts from the electric or magnetic field in order to emphasize
the effects from the AC Stark shift and the saturation of the
transition. Calculations were done for 0.1, 0.2, 0.5, and 1.0~W
of power in each beam.
The results are shown in
Fig.~\ref{fig:OBth} where we have suppressed the AC
Stark shift in (a) and (b) but shown the full results in
(c) and (d).

There are a few important trends that are worth noting. For the
calculations that suppressed AC Stark shift, Figs.~\ref{fig:OBth}(a)
and (b), only saturation of the 1S--2S is changing the results
from perturbation theory, Eq.~\eqref{eq:thrate}.
The decay of the atom while crossing the beam has a minor effect
for the parameters of these calculations. As foreshadowed above,
saturation is more important for slowly moving atoms which are the
ones that mainly contribute to the signal near zero detuning.
The region of detuning where the optical Bloch differs from the
perturbative results decreases with decreasing laser power.
Unsurprisingly, saturation is more important for the 10~mK
$\bar{\rm H}$s than for those at 50~mK due to the larger fraction
of slow atoms. The AC Stark shift in Figs.~\ref{fig:OBth}(c) and
(d) is the other important effect in these calculations. The size
of the shift is approximately the same for the two temperatures
because it depends on the path through the laser beam and not the
time in it. However, the size of the shift is a factor of $\sim 3.5$
smaller than the estimate from Eq.~\eqref{eq:ACst}. For example,
at 1~W,
the peak in Figs.~\ref{fig:OBth}(c) and
(d) are shifted by 1.5~kHz compared to 5.3~kHz
from Eq.~\eqref{eq:ACst}. The actual shift is smaller because
the 5.3~kHz is the shift at the intensity maximum whereas
the $\bar{\rm H}$s travel through the beam, experiencing both
large and small intensity, and they always miss
the exact center when they cross the beam so the peak intensity
on a particular crossing is less than the maximum.

\begin{figure}
\resizebox{86mm}{!}{\includegraphics{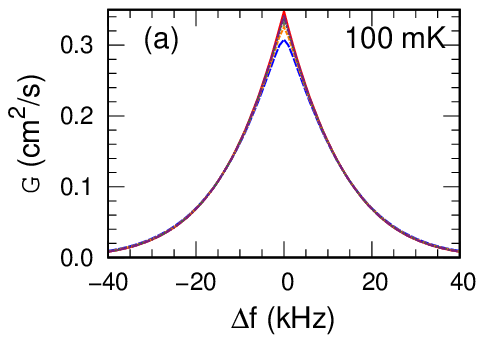}\includegraphics{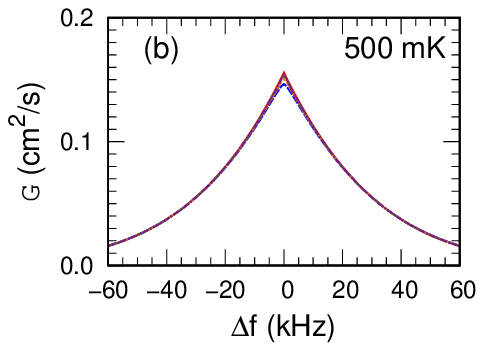}}
\resizebox{86mm}{!}{\includegraphics{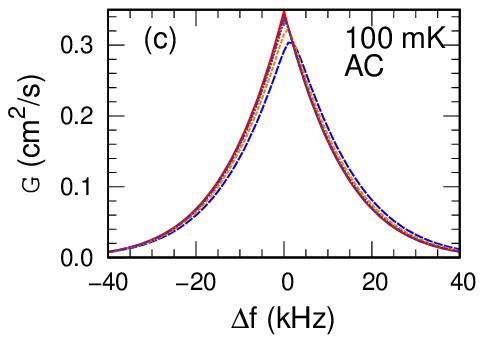}\includegraphics{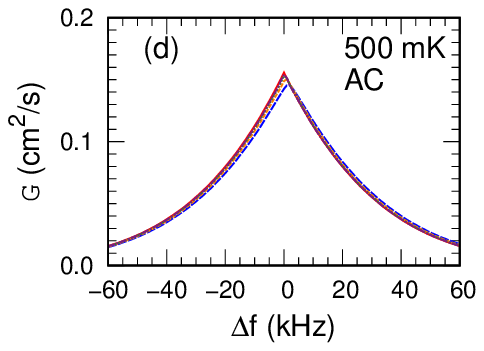}}
\caption{\label{fig:OBmax}
Same as Fig.~\ref{fig:OBth} except for a flat velocity distribution
up to the limit $|\vec{v}|=v_m$ with $(1/2)Mv_m^2/k_B = 100$~mK
for (a) and (c) and 500~mK for (b) and (d).
}
\end{figure}

\subsubsection{Energy dependent distribution}\label{sec:DistMax}

The results in this section are for the case where there is a
flat velocity distribution within a sphere in velocity space of
radius $v_m$
and zero otherwise. This gives ${\cal D}=(3/v_m^3)\sqrt{v_m^2-v_\perp^2}$
as discussed above. As with the previous section, we numerically
solve the optical Bloch equations to obtain $\rho_{11}$ and use
Eq.~\eqref{eq:OBrate} to calculate the rate.

These results are compared to the analytic, perturbative expression,
Eq.~\eqref{eq:hirate}, in Fig.~\ref{fig:OBmax}. For this case,
we have somewhat higher \Hb energies than the previous section
because previous experiments, Ref.~\cite{ALP2017,ALP2018}, have a trap depth
of $\sim 1/2$~K which matches Figs.~\ref{fig:OBmax}(b) and (d).
The shape of the rate versus frequency is qualitatively similar
to the previous results. There is a similar cusp feature for
the calculations that do not include the AC Stark shift,
Figs.~\ref{fig:OBmax}(a) and (b).
As with the previous section, the AC Stark shift,
Figs.~\ref{fig:OBmax}(c) and (d), gives a
$\approx 1.5$~kHz displacement of the peak position for 1~W
power in one beam.

The width for the 100~mK case in Fig.~\ref{fig:OBmax} has
approximately the same width as the 50~mK in Fig.~\ref{fig:OBth},
within 10\%. This is because the flat distribution within
a sphere is missing the higher energy $\bar{\rm H}$'s which
broaden the line. Although the line shapes are similar,
the thermal distribution falls faster at smaller detunings
and then slower at larger detuning reflecting the difference
in shape of a thermal and a flat distribution with respect
to speed. 

\section{Line shape: nonzero magnetic shift}\label{sec:LiShB}

In this section, we give results when the shifts in transition
frequency due to magnetic fields are included. The perturbative
transition rate can be calculated analytically for a thermal
distribution and power law potential
and reduced to a single integral for a distribution
which is equally likely for energy less than a limit. If the
distribution only depends on the energy, the perturbative
transition rate can be reduced to a two dimensional integral.

The main idea in this section is that the probability for crossing
the beam at a position, $z$, depends on the trapping fields
and will be represented by a probability distribution
of the $\bar{\rm H}$'s, ${\cal P}(z)$. The
shift in the transition frequency depends on the magnetic field
at the position $z$ as well. By convolving these effects with
the transition rate as a function of $z$, the overall transition
rate, ${\cal J}$, can be calculated. As with the previous section,
the ${\cal J}$ will have units of $area/time$.

If the $v_\perp$ distribution does not depend on $z$, then
the transition rate only has a $z$ dependence through the
transition frequency, $f_0(z)$: ${\cal G}(f_L-f_0(z))$.
There are distributions where the $v_\perp$ distribution
does depend on $z$ in which case we will indicate the extra
parametric dependence as ${\cal G}(f_L-f_0(z),z)$. An example of
this is a distribution which is flat in velocities and $z$
for $E<E_m$ and zero for $E>E_m$. The overall transition rate
is then
\begin{equation}\label{eq:TotRate}
{\cal J}=\int {\cal G}(f_L-f_0(z),z) {\cal P}(z) dz
\end{equation}
where the integral is over the region where ${\cal P}(z)$ is
nonzero.

\subsection{Perturbative result}

For weak lasers where the perturbative result is accurate, the
rate can be calculated from Eq.~\eqref{eq:avgrate} when given
the $v_\perp$ distribution, ${\cal D}(v_\perp )$, at the position
$z$. For the two special cases treated in the figures above, analytic
expressions, Eqs.~\eqref{eq:thrate} and \eqref{eq:hirate},
are available. For a general position distribution, ${\cal P}(z)$,
and magnetic field, $B(z)$,
the overall transition rate will result from a one-dimensional
integration, Eq.~\eqref{eq:TotRate}.

For the next two subsections, we assume that the magnetic field
has a simple form
\begin{equation}\label{eq:Bvsz}
B(z) = B_0+B_\nu z^\nu
\end{equation}
with $\nu$ an even integer to give an effective potential energy
that traps the \Hb in $z$. In Sec.~\ref{sec:CompB}, we discuss
more physical magnetic fields.

\subsubsection{Thermal distribution: power law potential and shift}

In this section, we assume the \Hb distribution is from a thermal
distribution in velocity and position.
In this case, the $v_\perp$ distribution
is independent of $z$ resulting in the transition rate in
Eq.~\eqref{eq:thrate}. The position distribution is
\begin{equation}
{\cal P}_{th}(z)=C e^{-\beta \mu B_\nu z^\nu}
\end{equation}
where $\mu$ is the magnetic dipole moment of the 1Sd state,
$C=\nu (\beta\mu B_\nu)^{1/\nu}/[2\Gamma (1/\nu )]$,
$\beta = 1/(k_B T)$ and $\Gamma (z)$ is the Gamma function.
Because $\mu$ will always be multiplied by $\beta$, which typically
won't be very well known, the magnetic dipole moment of the electron
can be used.
The constant in front of the exponential gives a normalized position
distribution.
In terms of $z$, the transition frequency versus $z$ can be found
from Eq.~\eqref{eq:dEdB} to give
\begin{equation}\label{eq:freqSh}
f_0(z) = f_0(0) + f_\nu z^\nu + f_{2\nu}z^{2\nu}
\end{equation}
where $f_0(0)$ is the frequency evaluated at $B=B_0$,
$f_\nu = (93.035 + 387.678 B_0) B_\nu$~kHz,
and $f_{2\nu }=193.839 B_\nu^2$~kHz where $B_0$ is in Tesla
and $B_\nu$ is in Tesla/meter$^\nu$ (to avoid the symbol $T$ which
could be confused with temperature). Putting together
with Eq.~\eqref{eq:thrate}, the overall transition rate is
\begin{equation}\label{eq:TotRateth}
{\cal J}_{th}=\frac{\pi A C}{2v_{th}}\int_{-\infty}^\infty w
e^{-|\Delta f-f_\nu z^\nu-f_{2\nu}z^{2\nu}|/\phi}e^{-\beta\mu B_\nu z^\nu}
dz
\end{equation}
where $\phi = v_{th}/(4\pi w)$, $\Delta f=f_L-f_0(0)$,
and the $z$-dependence of the waist, $w(z)$, is from Eq.~\eqref{eq:wgau}.
For the typical cases in the
ALPHA experiment, the waist is 200~$\mu$m giving $Z_R=0.52$~m.
The $z$-dependence in the waist will lead
to errors of $\sim$1\% in ${\cal J}$ and, therefore, we will
ignore this dependence. Even with this approximation, we have not
found an analytic expression for Eq.~\eqref{eq:TotRateth} and
evaluated it numerically.

For low temperatures, the \Hb can not reach magnetic fields substantially
larger than $B_0$. In this case, the contribution from $f_{2\nu}z^{2\nu}$
is negligible and the integral can be evaluated analytically.
For $f_{2\nu}=0$,
\begin{eqnarray}\label{eq:TotRatethapp}
{\cal J}_{th}(\Delta f<0)&=&\frac{\pi w_0A}{2v_{th}}\tau_+^{1/\nu}e^{\Delta f/\phi}\nonumber\\
{\cal J}_{th}(\Delta f>0)&=&\frac{\pi w_0A}{2v_{th}}\tau_+^{1/\nu}
\frac{\Gamma (1/\nu ,t_+)}{\Gamma (1/\nu )}e^{\Delta f/\phi}+\nonumber\\
&\null&\frac{\pi w_0A}{2v_{th}}\tau_-^{1/\nu}
\frac{\tilde{\gamma} (1/\nu ,t_-)}{\Gamma (1/\nu )}e^{-\Delta f/\phi}
\end{eqnarray}
where $\tau_\pm=\beta\mu B_\nu /[\beta\mu B_\nu \pm (f_\nu/\phi)]$,
$t_\pm = [\beta\mu B_\nu \pm (f_\nu/\phi)]z_0^\nu$, and
$f_\nu z_0^\nu =\Delta f=f_L-f_0(0)$ defines the position where the detuning is
zero. The incomplete gamma functions are defined as
\begin{equation}
\Gamma (s,x)=\int_x^\infty t^{s-1}e^{-t}dt    
\end{equation}
while the possibility for $t_-<0$ leads to the generalized definition
\begin{eqnarray}\label{eq:IncGam}
\tilde{\gamma} (1/\nu,t_- >0)&=&\int_0^{t_-} t^{(1/\nu )-1}e^{-t}dt
=\Gamma (1/\nu )-\Gamma (1/\nu ,t_-)\nonumber\\
\tilde{\gamma} (1/\nu,t_- <0)&=&\nu\int_0^{t_-^{1/\nu}}e^{-u^\nu}du\nonumber\\
&=&\nu t_-^{1/\nu }\sum_{n=0}^\infty \frac{(-t_-)^n}{n! (n\nu +1)}
\end{eqnarray}
where the $1/(\beta\mu B_\nu-(f_\nu/\phi))^{1/\nu}$ in
Eq.~\eqref{eq:TotRatethapp} from the
$\tau_-$ cancels the same term in the $t_-^{1/\nu}$ from
Eq.~\eqref{eq:IncGam} when $t_-<0$.

\subsubsection{Energy dependent distribution: power law potential and
shift}

In this section, we examine the case where there is a flat distribution in
$v_x,v_y,v_z,z$ with the condition that $E<E_m$. This is similar to the
condition in Sec.~\ref{sec:EnDepVel} but accounting for the potential
energy along $z$. For this case, we can use the result in Eq.~\eqref{eq:hirate}
that analytically includes the averaging over impact parameter and
$\vec{v}$
in the $z$-convolution, Eq.~\eqref{eq:TotRate}. In this case, the $v_m$
depends on $z$: $Mv_m^2(z)/2=E_m-\mu [B(z)-B_0]$. The probability
distribution is
\begin{equation}
{\cal P}_m(z)=v_m^3(z)\left[\int_{z_0}^{z_f}v_m^3(z')dz'\right]^{-1} \; {\rm for}\;
z_0<z<z_f
\end{equation}
where the $v_m(z_0)=v_m(z_f)=0$. Using the simplified $B(z)$ from
above gives
$z_f=-z_0=[E_m/(\mu B_\nu )]^{1/\nu }$ and
\begin{eqnarray}
{\cal P}_m(z)&=&\frac{1}{2z_f\; _2F_1(-\frac{3}{2},\frac{1}{\nu} ;1+\frac{1}{\nu} ;1)}\left[
1-(z/z_f)^\nu \right]^{3/2}\nonumber\\
&=&\frac{2\Gamma (\frac{5}{2}+\frac{1}{\nu})}{3\sqrt{\pi}\Gamma (1+\frac{1}{\nu} )z_f}
\left[
1-(z/z_f)^\nu \right]^{3/2}
\end{eqnarray}
Unfortunately, we have not found an analytic expression for the
convolution
\begin{equation}
\label{eq:TotRatehi}
{\cal J}_m=2\int_0^{z_f} {\cal G}_m(f_L-f_0(z),z){\cal P}_m(z) dz.
\end{equation}
However, the average rate can be easily evaluated numerically because it is a one
dimensional integral.

\subsection{Optical Bloch result}\label{sec:OBBf}

As with Sec.~\ref{sec:OB1}, the full calculation of the line shape
uses the ${\cal G}_{ob}$
from Eq.~\eqref{eq:OBrate} in the convolution of Eq.~\eqref{eq:TotRate}.
The AC Stark shift is included in all of the calculations in this
section. For all calculations, we use Eq.~\eqref{eq:Bvsz} with
$B_0 =0.1$ or 1~Tesla and $\nu=6$ with
$B_6=0.5/0.12^6$~Tesla/m$^6$. This gives a variation in $z$ very similar to
that in the ALPHA trap along the beam axis. To date, the ALPHA experiments
have $B_0\approx 1$~Tesla.

\subsubsection{Thermal distribution}

\begin{figure}
\resizebox{86mm}{!}{\includegraphics{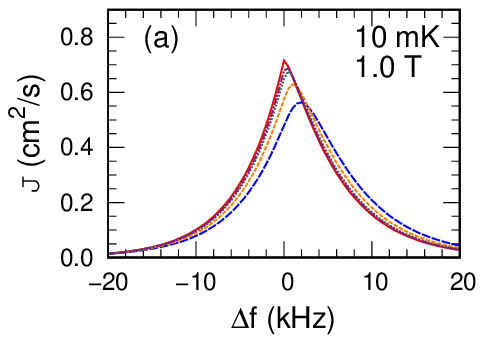}\includegraphics{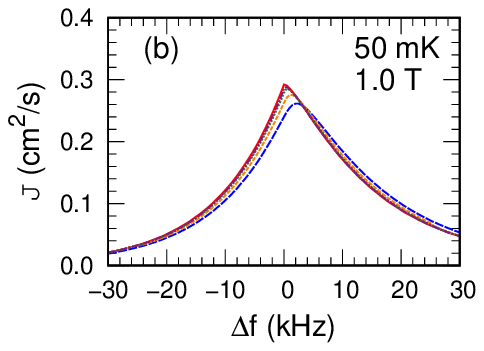}}
\resizebox{86mm}{!}{\includegraphics{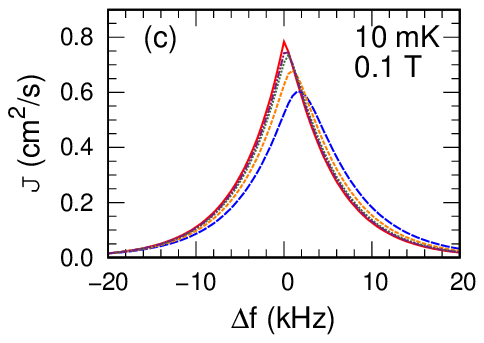}\includegraphics{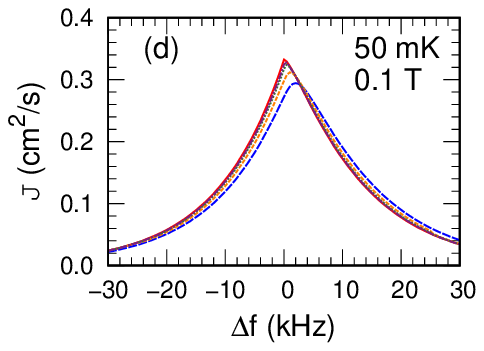}}
\caption{\label{fig:OBthZ}
Similar to Fig.~\ref{fig:OBth} except for Eq.~\eqref{eq:TotRate} which
includes the frequency shifts due to
a magnetic field of the form $B(z)=B_0+B_6 z^6$.
The red solid line does not include the AC Stark shift and was computed
using Eq.~\eqref{eq:TotRateth}. The other line types match the laser powers in
Fig.~\ref{fig:OBth}.
The $B_0=1.0$~Tesla are
plotted in (a) and (b) while the 0.1~Tesla are in (c) and (d). The $T=10$~mK
are in (a) and (c) while the 50~mK are in (b) and (d). All calculations
include the AC Stark shift.
}
\end{figure}

We present results, in Fig.~\ref{fig:OBthZ}, from numerically solving the optical
Bloch equations for different $\bar{\rm H}$ temperatures and
laser powers. In addition,
we changed the value of $B_0$ from 1.0~Tesla in Figs.~\ref{fig:OBthZ}(a)
and (b) to 0.1~Tesla in (c) and (d). These calculations illustrated a few
trends that will be important for future measurements.

The frequency shift from the magnetic field breaks the symmetry of the line
so that the decrease with positive detuning is slower than for negative
detuning. This effect increases with $B_0$ because the frequency shift
with magnetic field,
Eq.~\eqref{eq:dEdB}, increases with $B_0$. This leads to larger width
for larger $B_0$.
The increase of $d\Delta E/dB$ with $B$
is due to the diamagnetic term in the Hamiltonian for the 1Sd--2Sd
transition. The $d\Delta E/dB$
is roughly four times larger for 1.0~Tesla compared to 0.1~Tesla. Decreasing the
magnetic field further gives some decrease in $d\Delta E/dB$, but the
effect is not so large: only $\sim$30\% change going from 0.1~Tesla to
0~Tesla. 
As with the calculations in Fig.~\ref{fig:OBth}, saturation of the transition
and AC Stark shift plays an increasing role in going from 0.1~W to 1.0~W.
The saturation causes a suppression in the region of the peak which leads to
a more rounded maximum for the line shape at higher power. The AC Stark
shift moves the peak position by a somewhat larger amount, $\simeq 1.9$~kHz,
possibly due to the slower decrease for positive detuning.
Lastly, the $f_{2\nu}$ term in the frequency shift, Eq.~\eqref{eq:freqSh},
did not contribute a noticeable effect to the line shape because the
largest effect is for large $z$ where the detuning is large and the transition
rate varies slowly with $\Delta f$. Thus, the approximation
Eq.~\eqref{eq:TotRatethapp} works well for these cases at small power.

\subsubsection{Energy dependent distribution}

\begin{figure}
\resizebox{86mm}{!}{\includegraphics{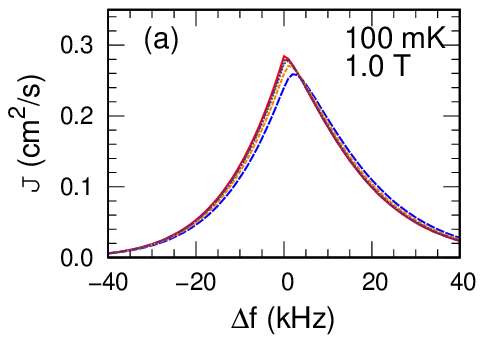}\includegraphics{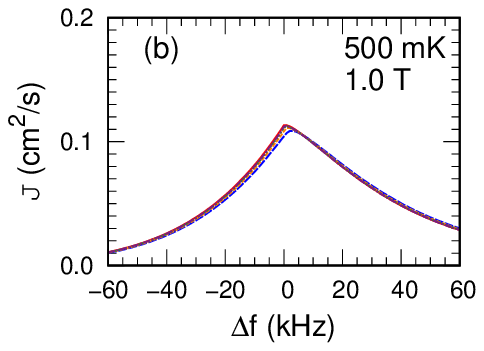}}
\resizebox{86mm}{!}{\includegraphics{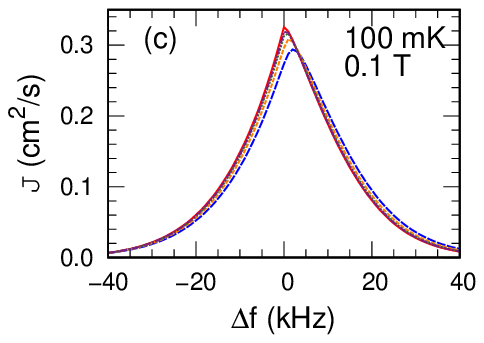}\includegraphics{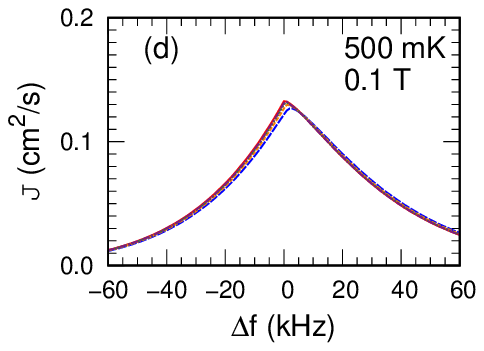}}
\caption{\label{fig:OBmaxZ}
Similar to Fig.~\ref{fig:OBmax} except for Eq.~\eqref{eq:TotRate} which
includes the frequency shifts due to
a magnetic field of the form $B(z)=B_0+B_6 z^6$.
The red solid line does not include the AC Stark shift and was computed
using Eqs.~\eqref{eq:TotRatehi}. The other line types match the powers in
Fig.~\ref{fig:OBth}.
The $B_0=1.0$~Tesla are
plotted in (a) and (b) while the 0.1~Tesla are in (c) and (d). The $E_m=100$~mK
are in (a) and (c) while the $E_m=500$~mK are in (b) and (d). All calculations
include the AC Stark shift.
}
\end{figure}

Similar to the previous section, 
we present results, in Fig.~\ref{fig:OBmaxZ}, from numerically solving
the optical Bloch equations for different $\bar{\rm H}$
cutoff energies and laser powers. In addition,
we changed the value of $B_0$ from 1.0~Tesla in Figs.~\ref{fig:OBmaxZ}(a)
and (b) to 0.1~Tesla in (c) and (d).

The results in this section hold similar lessons as the previous
sections.
For example, the magnetic field leads to an asymmetry
in the line with the asymmetry increasing with increasing $B_0$.
The AC Stark shift is somewhat larger than the case for no shift
with B-field: $\sim 2$~kHz for 1~W of power.
Also, a larger cutoff energy leads to a broader linewidth with the
effect somewhat smaller for a thermal distribution at the
temperature.

\subsubsection{More complex cases}\label{sec:CompB}

For the ALPHA experiment, the magnetic field does not have the simple
power law dependence of the previous sections. Thus, there aren't simple
analytic formulas that can be developed for ALPHA. However, the results
in the previous sections
point to the possibility that the extensive numerical simulations
used in previous studies\cite{ALP2017,ALP2018}
are not necessary. When three conditions are
satisfied, the line shape can be determined by integration: 1) the
distribution of trajectories is approximately known, 2) the laser
is sufficiently weak that AC Stark shifts and depletion of atoms
are negligible, and 3) the detection of $\bar{\rm H}$'s do not depend
on the frequency, $\Delta f$. The Eqs.~\eqref{eq:avgrate} and \eqref{eq:TotRate}
are used with the known spatial dependence of the magnetic field to obtain
the line shape. If the AC Stark shifts are non-negligible but
there is little depletion of atoms, then the optical Bloch equation
can be used so that Eqs.~\eqref{eq:OBrate} and \eqref{eq:TotRate}
will give the line shape.

In fact, the Fig.~\ref{fig:OBmaxZ}(b) is for similar parameters for the ALPHA
experiment.\cite{ALP2017,ALP2018} A comparison with the figures from
these papers shows a strong similarity with the 1~W example.

We carried out a calculation for a 50 mK
thermal distribution of $\bar{\rm H}$s
in the actual
ALPHA magnetic fields. We simulated their motion as in
Refs.~\cite{ALP2017,ALP2018} and their transition using the optical Bloch
equations. The only difference with the usual calculation was artificially
setting the detection efficiency to be independent of the $\bar{\rm H}$
position and velocity. We also assumed the  $\bar{\rm H}$ population
was not depleted which isn't the case in the experiments.
We compared this result to that using the ${\cal G}_{ob}$
from Eq.~\eqref{eq:OBrate} in the convolution of Eq.~\eqref{eq:TotRate}.
We found perfect agreement in this case. This comparison shows
our results can be extended to more complex magnetic fields.

\section{Optimum parameters}\label{sec:OptPa}

In this section, we discuss how various parameters affect the accuracy
of the 1Sd--2Sd frequency measurement. We will first address some of the
more obvious parameters (e.g. laser power and waist, uniform $B$-field
value, etc.) by discussing the trends in the line width. We will then
show that the predicted $\chi^2$ is useful for assessing less obvious
parameters (e.g. the number of frequencies and their values in a measurement).

To orient the discussion, note
the current uncertainty of the $\bar{\rm H}$ 1Sd--2Sd measurement is
at the few kHz level. Clearly, the immediate goal is to improve this
to the few 100 Hz level with a long term goal to reach the few Hz level.

{\it Laser power}:
In $\bar{\rm H}$ experiments, the AC Stark shift for 243~nm
laser at $\sim 1$~W is 1-2~kHz and is not currently the controlling factor in
the uncertainty. To reach uncertainties that are at the few 100 Hz level,
the laser power should be decreased by at least an order of magnitude since
the AC Stark shift is proportional to the laser power. Also important,
high laser power leads to a large fraction of the atoms transitioning
to the 2Sd state. Because the transition is detected by $\bar{\rm H}$
that are ejected from the trap, the characteristics of the $\bar{\rm H}$
population (i.e. position and velocity distribution)
changes when there is a large probability for a transition.
This is problematic because detailed modeling of the population
becomes necessary when a substantial fraction of the atoms are ejected.

{\it Laser waist}: The line width is mainly from the finite time for
an $\bar{\rm H}$ to cross the laser beam, i.e. transit 
broadening. By doubling the waist, this
contribution to the
line width will decrease by a factor of 2. This leads to
a more accurate determination of the transition frequency.

{\it $\bar{\rm H}$ temperature:} At lower temperature, the $\bar{\rm H}$
requires more time to cross the laser beam leading to smaller contribution
to the line width from transit broadening. Also, at lower temperature,
the $\bar{\rm H}$ can not reach as large $B$-field which also decreases
that contribution to the line width. Finally, shifts from the $v\times B$
effective electric field, are proportional to the temperature.
Laser cooling of $\bar{\rm H}$
has been demonstrated\cite{ALP2021}
as well as the effect on the line width. After laser
cooling, trap depths of $\sim 1/2$~K are not necessary. This would allow
for a controlled decrease in the depth of the trap. A slowly decreased
trap depth leads to adiabatic cooling which further improves the
measurements.

{\it Uniform $B$-field}: The size of $B_0$ affects the line width through
the diamagnetic term in the 1S and 2S energies. For larger $B_0$, the
change in frequency with changing $B$-field is larger which leads to a
larger line width. The current experiments typically occur with $B_0\sim 1$~T
because the plasmas used to make $\bar{\rm H}$ are colder, more stable,
and easier to diagnose at large $B_0$. It is possible to form $\bar{\rm H}$
at $\sim 1$~T and later ramp $B_0$ to lower values. The difficulty with
ramping the magnetic field is to precisely know the final value due to
persistent currents.
Thus, there is incentive to
keep $B_0\sim 1$~T. Fortunately, from Fig.~\ref{fig:OBthZ} and \ref{fig:OBmaxZ},
although there is a change in the line width in going from
$B_0=1$ to 0.1~T, the change is less than a factor of 2.
Thus, from the perspective of line width, there may not
be enough gained by decreasing $B_0$.

{\it $\chi^2$ treatment}: There are other
parameters that affect the accuracy of the measured 1Sd--2Sd transition
frequency but are not as obvious. For example, given 9 frequencies to
measure the transition, which frequencies should be chosen? Or, would it
be better to use 9 or 17 frequencies to measure the line? How does the
presence of background atoms affect the accuracy of the frequency
determination? In these cases,
we propose to use the $\chi^2$ of a calculated line shape to guide these
choices.

For this discussion, we will use the form in 
Eq.~\eqref{eq:TotRatethapp} with the parameters of Fig.~\ref{fig:OBthZ}(b)
as the exact transition and will briefly investigate the role played by the
frequencies chosen in the measurement. We will have $n_f$ frequencies with
$\Delta f_j$. We will also include the possibility that all transitions
are shifted by $\delta f$. We can compute a synthetic line by using
Monte Carlo to randomly determine the number of atoms, $N_j(\delta f)$,
to make a transition at frequency $\Delta f_j$
for a fixed total number $N$.
On average, this leads to $\bar{N}_j(\delta f)$ atoms making the transition
with
\begin{equation}
\bar{N}_j(\delta f)=N{\cal J}_{th}(\Delta f_j+\delta f)/\sum_j
{\cal J}_{th}(\Delta f_j+\delta f)
\end{equation}
with $N$ the total number of atoms making the transition.
We computed a $\chi^2$ by averaging over many different realizations
of the Monte Carlo simulation
\begin{eqnarray}
\chi^2&=&\langle\sum_j [\bar{N}_j(0)-N_j(\delta f)]^2\rangle /\bar{N}_j(0)
\nonumber\\
&\simeq &n_f-1+\sum_j [\bar{N}_j(0)-\bar{N}_j(\delta f)]^2 /\bar{N}_j(0)
\end{eqnarray}
where the $\langle ...\rangle$ on the first line means to average over
the different realizations and the second line is from Poisson
statistics. Because we fix $N$, the number of degrees of freedom
is $n_f-1$.

\begin{figure}
\resizebox{86mm}{!}{\includegraphics{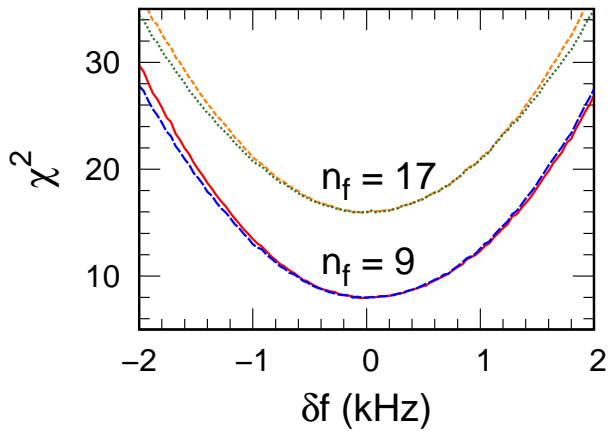}}
\caption{\label{fig:chi2}
Calculations with 9 frequencies are (red solid) for spacing 1
and (blue dashed) for spacing 1/2. Calculations with 17 frequencies
are (orange short dash) for spacing 1 and (green dotted) for spacing 1/2.
All calculations were for $N=1000$ transitions.}
\end{figure}

In Fig.~\ref{fig:chi2}, we use the $\chi^2$ to see how choices
for the frequencies can affect the accuracy for which the transition
is determined. Instead of allowing all detunings to be freely varied,
we started with a symmetric choice similar to that used in an ALPHA
experiment. We did four choices for the frequencies. Spacing
1 for $n_f=9$ were the frequencies $\Delta f_j
=0,\pm 5,\pm 10,\pm 20,\pm 50$~kHz
while spacing 1/2 divided every frequency by 1/2. Spacing 1 and 1/2 for
$n_f=17$ also used the frequencies halfway between those
for $n_f=9$, i.e. $\Delta f_j = 0, \pm 2.5, \pm 5, \pm 7.5,...$~kHz.
For 8 degrees of freedom, $\chi^2\simeq 20$ corresponds to a p-value of
0.01 while this corresponds to $\chi^2\simeq 32$ for 16 degrees
of freedom. Visually, it is clear that the spacing 1 calculations give
slightly greater curvature and therefore modestly
better bounds on the uncertainty in
the frequency. To compare 9 versus 17 points,
the p-value of $\simeq 0.01$ corresponds to
$\delta f\simeq -1.5,1.6$ for $n_f=9$ and $\delta f\simeq -1.8,1.8$
for $n_f=17$ which means the $n_f=9$ will give a somewhat better bound
on the transition frequency. More importantly, this suggests using
$\chi^2$ as a metric for choosing the number and values
for the frequency.

As a simple extension, we use the ideas of this section to estimate
parameters needed to get to few 100~Hz accuracy. If the laser waist
is increased from 200 to 400~$\mu$m, the uncertainty decreases by a
factor of 2. Increasing the number of detected $\bar{\rm H}$'s
from 1,000 to 4,000 decreases the uncertainty by another factor of 2.
These two improvements with the estimate from the previous paragraph
leads to a few 100~Hz accuracy.

This example is somewhat artificial because the $\bar{\rm H}$
temperature may not be well known even if the distribution is approximately
thermal. In this case, the $\chi^2$ can be calculated versus
$T$ and $\delta f$. We have done this for the $n_f=9$, spacing 1 case
and found that the $\chi^2$ gave $30<T<80$~mK (compared to 50~mK of
the actual distribution) with the range of
$\delta f$ similar to that found above. This showed that a simultaneous
fit could give reasonable results. The example is also artificial
in that we did not include the effect of frequency independent
background atoms; the background atoms will somewhat increase the
uncertainty in the frequency but should not skew the results.
Finally, in the real experiment, the detection efficiency could
depend on the frequency of excitation, $\Delta f_j$, which would
skew the results. We have not treated the change in line shape due
to detection efficiency but it can be incorporated into our treatment
if it is known. These artificial conditions can be easily removed
in the numerical implementation of the $\chi^2$ method. We have not
done so here because they depend on specific aspects of future experiments.

\section{Summary}\label{sec:Sum}

We have examined some of the physics that determines the line shape of the
1Sd--2Sd transition in magnetically trapped $\bar{\rm H}$. Under three
assumptions (the
distribution of trajectories is approximately known, the laser
is sufficiently weak that AC Stark shifts and depletion of atoms
are negligible, and the detection of $\bar{\rm H}$'s do not depend
on the frequency, $\Delta f$), the line shape can be calculated as
an integral over a few degrees of freedom. If the AC Stark shift is
not negligible, solutions from optical Bloch equations can be used
in these integrals. In either case, large scale, detailed simulations
of the trajectories would not be needed to obtain the line shape.
We presented analytic expressions for the transition rates for
special cases of the $\bar{\rm H}$ distribution and magnetic field.

We discussed several of the trends that control the accuracy with which
the transition frequency can be determined. These include parameters such as
$\bar{\rm H}$ distribution, laser waist and intensity, uniform $B$-field,
number and choice of frequencies sampled, and others. We also propose
the use of a $\chi^2$ test to optimize the choices for these parameters.
From the discussions above, it seems that modest improvements in the
ALPHA experiment could increase the accuracy of the 1S--2S transition
by an order of magnitude.

Further exploration is needed to project
the best path to reach accuracy comparable to that in experiments
on normal matter H. Table 3 of Ref.~\cite{ALP2018} gives the sizes of
various sources of uncertainties in the 1S--2S transition frequency.
Statistical uncertainties (Poisson errors and curve fitting) and modeling
uncertainties were the largest sources at 3.8 and 3 kHz
respectively; these were addressed above.
The next largest uncertainty was laser frequency stability at 2~kHz; this
can be decreased to the several Hz level
by using a different stabilization method. The
next largest uncertainty was the absolute magnetic field measurement
at 0.6~kHz; this can be decreased by decreasing the size of the magnetic field,
Eq.~\eqref{eq:dEdB}, or through a more accurate determination of $B$.
The next largest uncertainty is from the discrete choice of frequencies
at 0.36~kHz; this was addressed in Sec.~\ref{sec:OptPa}. The next largest uncertainties
were DC-Stark shift, at 0.15~kHz, and second order Doppler shift at 0.08~kHz;
these can be decreased by using colder $\bar{\rm H}$s since they both
are proportional to the kinetic energy, in fact, using Ref.~\cite{ALP2021}
we estimate these will decrease by a factor of $\sim 10$ with already demonstrated laser
cooling. Of these, the most problematic uncertainty could be from Poisson errors
because it will require a couple order of magnitude increase in the
number of $\bar{\rm H}$s to decrease the Poisson errors to the several
Hz level.

Data used in this publication is available at~\cite{data}.

This work was supported by NSF grant PHY-1806380.

\bibliography{HbarLineshape}

\end{document}